\documentclass[aps,pra,superscriptaddress,twocolumn,longbibliography]{revtex4-2}

\usepackage{helvet}

\usepackage{soul}

\usepackage{palatino}
\usepackage[utf8]{inputenc}
\usepackage{epsf}
\usepackage{amsmath,amssymb}
\usepackage{calc}
\usepackage{amsmath,amssymb, bm}
\usepackage{float}

\usepackage[usenames,dvipsnames]{xcolor}

\usepackage{graphicx}
\usepackage{bm}
\usepackage[normalem]{ulem}
\usepackage{amsbsy,amsmath}

\usepackage{hyperref}

\newcommand{\ben}{\begin{equation}}
\newcommand{\een}{\end{equation}}
\newcommand{\bea}{\begin{eqnarray}}
\newcommand{\eea}{\end{eqnarray}}

\def\sss{\scriptscriptstyle\rm}

\def\1s{_{1,\sss S}}
\def\2s{_{2,\sss S}}

\def\s{_{\sss S}}
\def\xc{_{\sss XC}}

\def\H{_{\sss H}}

\def\ext{_{\rm ext}}

\def\br{{\bf r}}

\begin{document}
\title{Dysfunctional Functionals and their Therapies in Time-Dependent Density Functional Theory}
\author{Neepa T. Maitra}
\affiliation{Department of Physics, Rutgers University at Newark, 101 Warren Street, Newark, NJ 07102, USA}
\email{neepa.maitra@rutgers.edu}
\date{\today}

\begin{abstract}
The exchange-correlation functional in  Time-Dependent Density Functional Theory (TDDFT) is known to have dependence on past densities and initial states, and yet this memory-dependence is ignored in practical calculations which almost always use an adiabatic approximation. We discuss the dysfunction that arises due to this deficiency and point to therapies that go beyond the adiabatic approximation. Still, adiabatic TDDFT has had success in a wide range of systems for response properties and for fully non-equilibrium dynamics, and why this is so is only partially understood. This, together with 
the  rigorous foundation of the exact theory and its favorable system-size scaling,  provides tremendous impetus for continuing to strive towards improved approximations to the exact functionals, and a deeper understanding of the impact of approximations in applications.
\end{abstract}

\maketitle

\section{Introduction}
A century after the publication of the Schr\"odinger equation, we may step back and marvel at the fundamental understanding and technological developments that the mathematical formulation of non-relativistic quantum mechanics has provided, with novel subfields and applications continuing to emerge today. In practise, the equation would be far less impactful without the formulation of reduced-variable theories: the description of even the smallest relevant systems typically involve many electrons and many nuclei that solving Schr\"odinger's equation not only pushes computational limits but also the many-variabled wavefunction gives us far more information than what we could possibly want. Considering the description of the electronic subsystem, density functional methods are the most widely-used, due to their balance between accuracy and efficiency while resting on a rigorous foundation. 
For matter interacting with light or otherwise disturbed from equilibrium, time-dependent density functional theory (TDDFT) was formulated almost 40 years ago~\cite{RG84,Carstenbook,GM12chap}, reviewed in more than ten reviews in the past ten years~\cite{M16,PI16,GLL18, LGIDL20,S21,YCLZM21,SBLJ21,Herbert23,LM23,XCZCY24,SHGR25,Amrutkar2025,U25}. TDDFT is  based on proving a one-to-one map between the time-evolving one-body density of a many-body system and the external potential, for a fixed initial wavefunction.

The proof itself is both a beauty and a beast: the beauty is that it is based on Heisenberg's equation of motion for the current-density, making it relatively straightforward to understand after even one course in quantum mechanics. However, a beast hides inside because of implied time-analyticity properties that do not typically hold in the presence of spatial singularities such as the Coulomb potential. This has triggered the development of subsequent proofs attempting to avoid this requirement~\cite{MTWB10,FLLS16,RL11,RPL15,LLO26}; most notably the recent work of Ref.~\cite{LLO26}.

TDDFT operates by solving a set of coupled single-particle non-linear Schr\"odinger equations, mimicking the interacting electronic system by a non-interacting Kohn-Sham (KS) one, with all the quantum many-body effects hiding in the exchange-correlation (xc) potential, $v\xc[n;\Psi(0),\Phi(0)](\br,t)$. At time $t$ and position $\br$, this potential is a functional of the one-body density $n(\br',t') = N\sum_{\sigma',\sigma_2...\sigma_N}\int d^3r_2...d^3r_N \vert\Psi(\br'\sigma',\br_2\sigma_2 ... \br_N\sigma_N, t)\vert^2 $ at all points in space $\br'$ and all past times $t' \le t$ , the initial many-body interacting state $\Psi(0)$ and the initial KS state $\Phi(0)$. The theorems state that with the use of the exact functionals, all observables of the true interacting electronic system  can in principle be extracted from that of the KS system~\cite{RG84,Carstenbook,GM12chap,M16,U25}.
This dependence on the past ({\it memory-dependence}) is a price to pay for the reduced description, as is the non-linearity;  the many-body Schr\"odinger equation is linear, and needs only the instantaneous wavefunction in order to propagate forward. 

While the non-linearity itself has not appeared to pose any numerical problem, practical approximations that contain appropriate memory-dependence remain elusive. The usual approach is to completely neglect memory and use the xc potential for a {\it ground state} of the same instantaneous density,  $v\xc^A[n;\Psi(0),\Phi(0)](\br,t) = v\xc^{\rm g.s.}[n(t)](\br,t)$, known as the {\it adiabatic approximation.} 
In this Perspective, we revisit both the success and the dysfunction that can arise from such a functional approximation, and the therapies that have been put forward to go beyond the adiabatic approximation.

\section{Pleasure and Pain: Successes and Failures of Adiabatic TDDFT}
\begin{figure}[h]
\centering
\includegraphics[width=\linewidth]{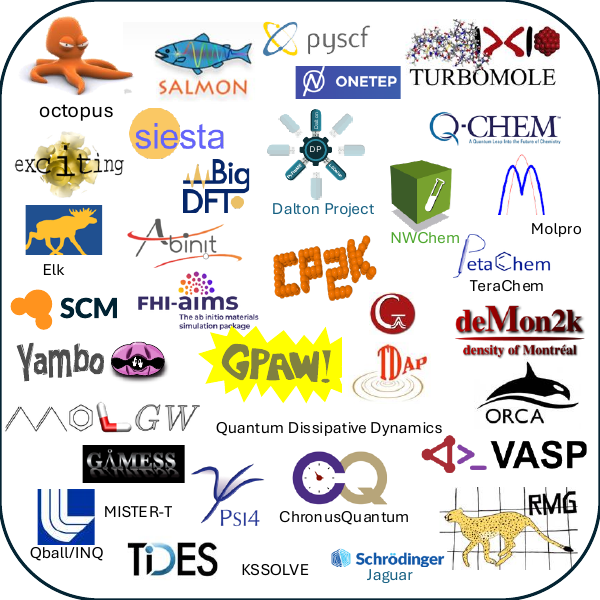}
    \caption{Codes with TDDFT capability, either real-time TDKS propagation, or linear response, or both~\cite{octopus2, salmon, pyscf, turbomole, onetep, exciting, siesta, nwchem,dalton, qchem, elk, molpro, abinit, cp2k, gaussian16, deMon2k, ADF, fhi-aims, gpaw, tdap,orca, chronusq, molgw, Qball-INQ,DFTBp,yambo,QDD,gamess,bigdft,psi4,jaguar,TiDES,KSSOLVE, RMG,vasp,mister-t,terachem} have widened the range of scientists who can run TDDFT calculations, providing truly invaluable tools for both experimentalists and coding-shy theorists.
    }
    \label{fig:codes}
\end{figure}

Adiabatic TDDFT is now a central part of many codes (see Fig.~\ref{fig:codes}) where it has enabled the quantum mechanical study of complex and fascinating phenomena, that would be computationally impossible otherwise. 
In what is often called real-time TDDFT,  the KS equations are propagated in real-time (TDKS):
\ben
(-\nabla^2/2 + v\s(\br,t))\phi_i(\br,t) = i \partial_t\phi_i(\br,t)
\label{eq:tdks}
\een
where 
$
v\s(\br,t) = v\ext^{(0)}(\br) + v\ext^{\rm app}(\br,t) + v\H[n](\br,t) + v\xc[n;\Psi(0),\Phi(0)](\br,t)
$
is the sum of the static external potential due to nuclei, an applied, possibly time-dependent, potential such as a laser field, the Hartree potential representing the classical Coulomb potential of the density, and xc potential, typically approximated by an adiabatic approximation. The resulting instantaneous dependence of the Hamiltonian on the density $n(\br,t) = \sum_i^N\vert\phi_i(\br,t)\vert^2$ simplifies the computational implementation of the equations compared to if a memory-dependent functional was used, allowing similar propagators to be used as for the linear TDSE, with the non-linearity entering through predictor-corrector approach for example~\cite{GMRC18,ZH18,GBC20}. 
In the limit of weak perturbations around a ground state, usually Eq.~\ref{eq:tdks} is instead cast in frequency-space through response equations derived from time-dependent perturbation theory. Linear response gives access to the excitation energies and properties related to the ground-to-excited transition densities of the system (e.g. oscillator strengths)~\cite{PGG96,C95,BA96a}. Excited-state densities are available from considering time-independent perturbation theory of the excitation energies computed from linear response~\cite{FA02,F01,BM25}, which could equivalently be framed as a variational problem. Quadratic response has been formulated for properties related to transition-densities between excited states (e.g. as needed for their non-adiabatic couplings, hyperpolarizabilities, two-photon absorption)~\cite{SS87,GDP96,SVHA02,PF18}. 

In the following, we will discuss how the adiabatic approximation gives both pleasure and pain in these two regimes, considering first linear response and then non-perturbative.

\subsection{Linear response regime}
By far, the dominant applications have been in the linear response regime, to compute the optical spectra of molecules and materials, as well as in the computation of potential energy surfaces and their couplings in photochemical applications~\cite{PGG96,C95}. Simulations treating an unprecedented number of electrons and excitations explicitly have been enabled, e.g. the green fluorescent protein complex of 4353 atoms with 40,518 atomic orbitals in Ref.~\cite{Kim2023}. High-energy-density materials have been studied, without the need for arbitrary partitionings~\cite{KNB26}; for large systems, spectra and response properties can arguably be obtained most efficiently by real-time propagation under a weak perturbation, and the same propagation framework can also be used with stronger fields.
Quadratic response TDDFT for molecules  for excited state absorption has been recently benchmarked~\cite{SGDLJ25}.

Turning to the role of memory in these calculations, let us consider first {\it where} the xc functional enters these calculations. The xc functional approximation involves $v\xc[n; \Psi(0),\Phi(0)](\br,t)$ in a domain where the time-dependence in $v\ext^{\rm app}(\br,t)$ is weak such that only first-order responses to it are considered, and the initial states $\Psi(0),\Phi(0)$ are respectively the interacting and KS ground-state wavefunctions of $v\ext^{(0)}(\br)$ and the KS counterpart $v\s^{(0)}(\br)$ (and by Hohenberg-Kohn theorem~\cite{HK64} no explicit initial-state dependence is needed here). 
While this implies a single choice is needed for the approximation for $v\xc[n](\br,t)$,
it is important to realize that in practise the xc functional usually enters in two distinct places in response calculations. That is, in practise, one separates the ground-state calculation from the perturbative one, so that the xc functional appears first in the choice of the ground-state functional out of which the KS response is calculated (e.g. molecular orbitals and their orbital energies, or in the case of extended systems, the density-density response function of the KS ground state), and second in the dynamical response, embodied in the functional derivative in frequency-space $f\xc[n^{\rm g.s.}](\br,\br',\omega)$. Formally, the two choices ($v\xc^{\rm g.s.}[n^{\rm g.s.}]$ and $f\xc[n^{\rm g.s.}]$) should be consistent with a single choice for the perturbative-domain $v\xc$, but often they are considered separately. This can be for  practical reasons: For example, one might take a higher-level approximation for $v\xc^{\rm g.s.}[n^{\rm g.s.}]$, than for the dynamical part $f\xc$, especially if the sophistication in the former is such that its functional derivative has not been computed or implemented.
For semiconductors and insulators, xc kernels  for excitonic spectra are more often than not developed independently of the ground-state approximation~\cite{BSU20,BSSR07}, and the band-gap is often evaluated as a ground-state problem~\cite{BAHTMB19}, needing a derivative discontinuity or an orbital functional treated with generalized KS, but it could equally be viewed as an excitation property~\cite{GM12chap}.  

For the question of memory-dependence, it is certainly useful to think of the ground-state and response parts separately, and away from $\omega=0$ there would be no conflict in these arising from an underlying $v\xc[n](\br,t)$. Time-non-locality translates to $\omega$-dependence in $f\xc(\br,\br',\omega)$, and ignoring this dependence, as an adiabatic approximation does,  places limitations on the accuracy when multiple excitations are involved as we will now discuss. 

\begin{figure}
\centering
    \includegraphics[width=\linewidth]{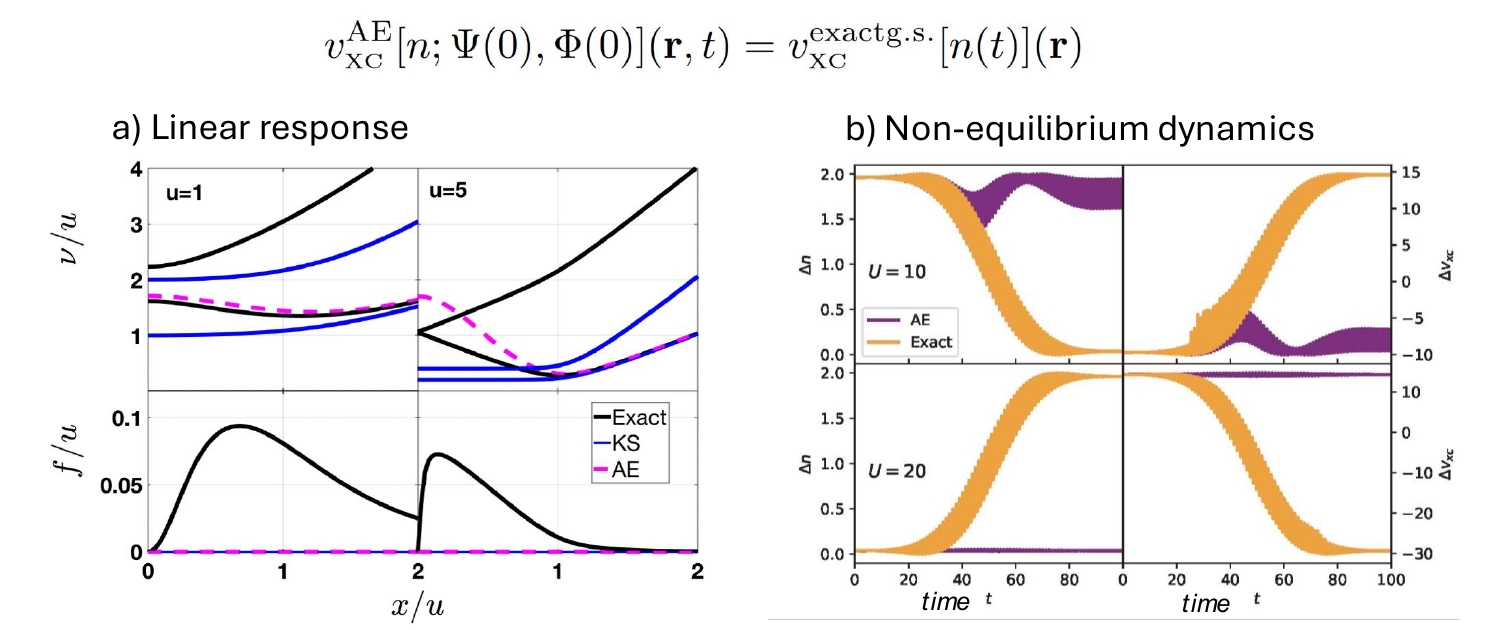}
\caption{Dysfunction from neglect of memory-dependence in the adiabatic approximation for the asymmetric Hubbard dimer, even when the exact ground-state functional is used (adiabatically-exact (AE)). 
\newline
{\bf a)} Linear response regime: Excitation frequencies $\nu$ (top panel) and a measure of relative oscillator strength $f$ for the second excitation (lower panel), in units of $u$, for dimensionless Hubbard parameters $u = U/2t = 1$ (left) and $5$  (right), as a function of scaled site potential difference $x/u = \Delta v/u$, with exact (black), KS (blue), and AE (pink)~\cite{CFMB18}. Note the scaled site potential scans over the Mott-Hubbard regime $x/u<1$ to the charge-transfer regime $x/u> 1$, and all values shown for $x$ in the case of $u=1$ and for $x/u > 1$ for $u=5$ are characterized as weakly-correlated regimes~\cite{CFSB15,CFMB18}. The AE excitation energy provides an excellent approximation to the true lowest excitation energy in these weakly-correlated regimes, providing a useful correction of the bare KS value. However the AE completely misses the second excitation over the entire range of parameters, and likewise its relative oscillator strength, due to the inability of AE to capture contributions due to double excitations. It also performs poorly in the strongly-correlated regime ($x/u <1, u=5$) for either excitation, since static correlation means that the excitations have a significant contribution from the double-excitation. 
While the adiabatic approximation is good for states dominated by single-excitation character, memory-dependence included in a frequency-dependent kernel is necessary to capture double-excitation character~\cite{TAHR99,TH00} and 'dressed' TDDFT approximations have been developed for this purpose~\cite{MZCB04,CZMB04,C05,MW09,MMWA11,RSBS09,SROM11,HIRC11,CH12,CH16,M22,DM23,DM25,DM25Err,Gao15}.
\newline
{\bf b)} Non-equilibrium dynamics: 
Site occupation density difference (left) and xc potential difference (right), for dynamics under a $\pi$-pulse resonant with the lowest excitation as a function of time,  for parameters $t = 1/2, \delta v\ext = -15$ exact (orange), and AE (purple). The top panel, with $U = 10$ is in the charge-transfer regime, while the bottom panel with $U = 20$ is in the Mott-Hubbard regime.
While the system remains close to the ground-state in the initial stages, the adiabatic approximation works well but as soon as it the system begins to finitely populate the excited state, the adiabatic approximation fails dramatically in all cases. The exact xc potential develops a stable step feature as time evolves in both cases, that serves to hold the transferred charge in place, in addition to oscillating steps (and in the real-space case, peaks), that are completely missed by any adiabatic approximation~\cite{EFRM12,FERM13,FM14b,RG12,M17}. 
While practical non-adiabatic solutions exist for $f\xc[n^{\rm g.s.}]$ in the response regime, developing approximations for non-adiabatic $v\xc[n; \Psi(0),\Phi(0)]$ remain a challenge. Ref.~\cite{DBM24,RBDM26} offers a way out through a formulation that uses response xc quantities to simulate non-equilibrium dynamics. 
}
\label{fig:hubbard}
\end{figure}

Linear response suggests single-photon processes, which in a non-interacting picture imply a response that involves only single excitations. Correspondingly, the KS response function is composed of single transitions from an occupied to an unoccupied orbital of the KS system. 
On the other hand, excitations of the true interacting system typically involve linear combinations of single, double, and higher excitations, relative to some non-interacting reference expansion,  and moreover the correlated nature of the ground-state means one cannot clearly delineate `occupied' and `unoccupied' orbitals in the first place, except in the weak-correlation limit~\cite{M22}. When applying the TDDFT linear response formalism to the non-interacting KS system, the xc kernel $f\xc$ mixes contributions from the KS single excitations, but any adiabatic approximation to it is blind to double and higher multiple excitations~\cite{MZCB04,CZMB04,C05,MW09,MMWA11,RSBS09,SROM11,HIRC11,CH12,CH16,M22,DM23,DM25,DM25Err,Gao15}. Panel a) of Fig.~\ref{fig:hubbard}, and the discussion in the caption, compare the true, exact KS, and adiabatically-exact TDDFT spectrum for an exactly-solvable model system, showing the consequences of failing to account for double-excitation character. 

States of double excitation character appear even in lower parts of an excitation spectrum however~\cite{KBLJ24}, and their propensity only increases when considering a full landscape of nuclear geometries encountered during photo-induced dynamics~\cite{LKQM06}. In extended systems, while plasmons are typically well-captured in even the simplest adiabatic approximation, ALDA, double- and higher plasmons are simply absent~\cite{Gao15,PGR18}. The lack of double-excitations is likely responsible for the poor predictions of thermally activated delayed fluorescence, predicting incorrect signs of singlet-triplet gaps~\cite{deSilva19}. For strongly correlated systems, where a single-determinant ground state description is qualitatively inadequate for the true ground state, double excitations in the KS system are important for every excitation~\cite{M22}: even if an extremely accurate ground-state functional was found such that the KS single-determinant ground state yields an accurate ground state density, static correlation means that all excitations out of the KS ground state would be near a double excitation where a second electron is excited to a near-degenerate orbital~\cite{M22}. With usual ground state approximations, the ground-state description is itself very poor in these situations, which plagues conical intersections involving them~\cite{LKQM06,TTRF08,TTC04}.

There are other instances where the lack of frequency-dependence in the xc kernel is responsible for significant errors, which could be interpreted in terms of missing multiple excitations but are often not discussed in that context because of their nesting in continuous spectra: the excitonic Rydberg series in solids~\cite{GZKT23}, the opening of the KS band-gap in semiconductors~\cite{GVbook,GM12chap,CBR20}, and dissipation in extended systems~\cite{AV06,AU06}. Frequency-dependence is also how the exact kernels overcome unphysical divergences in quadratic response TDDFT ~\cite{PRF16,LL14,OBFS15,ZH15,DRM23}.

Not all the errors in the linear response regime can be attributed to the adiabatic approximation however; sometimes the spatial-dependence is at fault. We refer the reader to a short discussion in Sec.~\ref{sec:therapies} and to the recent reviews cited in the introduction for a more complete  details of the pleasures and pains of linear response TDDFT. 

\subsection{Fully non-perturbative regime}
Going beyond linear response and allowing strong fields reveals a wealth of physics and phenomena: attosecond-scale time-resolved dynamics of molecules and materials, charge migration, dielectric breakdown in insulators, radiation chemistry, stopping power of projectiles,  high-harmonic generation, ultrafast demagnetization and optically-induced interlattice spin transfer,  Floquet band engineering, and its optimal control, biomimetic photovoltaic candidates; we refer the reader to the reviews over the past ten years and references therein~\cite{M16,PI16,GLL18, LGIDL20,S21,YCLZM21,SBLJ21,Herbert23,LM23,XCZCY24,SHGR25,Amrutkar2025,U25}.
 Again TDDFT allows simulations at unprecedented scales, e.g. Ref.~\cite{DAGBSC17} studied conductivity in a disordered aluminium system that treated almost 60,000 electrons explicitly. 
Adopting a term from Ref.~\cite{RCFB08}, it is fair to say that TDDFT has provided usefully accurate input in these applications, illuminating physical mechanisms and interpretations.

But, stepping back, one might wonder why do such calculations work?! 
This is not well-understood; the performance of the adiabatic approximation is much better characterized in the response regime. Various reasons have been pondered but none are entirely satisfactory. For example, could it be that in many of these applications $v\ext^{\rm app}(\br,t)$ dominates and one begins in a ground-state, so that the role of xc is really only needed to compensate self-interaction in  $v\H(\br,t)$, which adiabatic functionals approximately do to varying degrees of effectiveness? Even if  this is the case sometimes, it should also not be read to mean that an adiabatic self-interaction-free functional would give problem-free results e.g. Ref.~\cite{DRM26}. 
Could it instead be that the observable of interest is a relatively insensitive quantity, and the experimental reference is not so well-resolved? Most observables are spatially-integrated; sub-Angstrom spatial resolution in time-resolved experiments on the electronic time-scale exist but are challenging. 
Could it be due to a judicious choice of initial KS state in cases where the simulation does not begin in a ground state? That is, if the KS initial state has a similar configuration to that of the excited state, then one might be able to argue that remaining correlation could be similar to that of a ground state, and so the adiabatic approximation might provide reasonable results even for this far-from-ground state situation. Again this is not entirely satisfactory since different initial configurations reproduce the same density for excited or non-stationary states and have vastly different exact xc potentials even though the adiabatic ones are identical~\cite{FNRM16,EM12}; further, the configuration changes significantly over time as reflected by the rank of the interacting one-body density matrix while that of the KS system must remain constant, and so any advantage from a judicious pick of the KS initial state would be lost over time~\cite{LM20b}.
Finally, could it be that, through its lack of any memory-dependence, the adiabatic approximation ironically satisfies in a zeroth order way exact conditions (see Sec.~\ref{sec:therapies}) to do with memory? 

In fact, adiabatic TDDFT does not always yield usefully accurate results in the non-equilibrium regime; disagreements with experiment make it into published papers less often than agreements in successful applications, which is unfortunate as it would help us to understand why it does work when it does.
Sometimes, more sophisticated functionals perform worse than simpler ones~\cite{NTR24}.
Some failures which lead to unphysical results are well-documented and understood. For example, 
resonantly-driven dynamics targeting a single-excitation of a closed-shell system~\cite{RB09,EFRM12} cannot be captured. 
At the root of the problem is the inherently non-ground-state nature of the evolving state: xc correlation that pretends such a state is a ground state does not capture step and peak features of the exact xc potential, which depend on the history  of the density and the initial states~\cite{EFRM12,DLM22}.
It is perhaps not surprising such dynamics is so troublesome for TDKS when one thinks what the non-interacting KS system is trying to do: a minimal description of a single excitation  involves two orbitals, while the KS system starting in doubly-occupied orbitals must remain with doubly-occupied orbitals for all time due to the one-body nature of the KS potential and so, one orbital must describe at the same time both an electron that stays and one that gets excited. 

A striking example is laser-driven charge-transfer dynamics. Although optimally-tuned range-separated hybrids~\cite{BLS10,KSRB12, SKB09,KKK13,KB14} provide a reliable tool to obtain excitation energies of long-range charge transfer nature from TDDFT, applying these functionals, or any adiabatic functional, to simulate laser-driven dynamics from the ground to the charge-transfer excited state generally fails~\cite{RN11,FLSM15, FERM13,M17,DBM24}. Studies of model systems show that, when the charge-transfer occurs out of a closed-shell fragment, not only non-adiabatic step and peak structures evolve in time in the exact xc potential, but also a gradual step feature develops in time that stabilizes the charge-transfer state~\cite{FERM13,M17}. The latter is completely missed by most adiabatic functionals and only partially captured with the adiabatically-exact, $v\xc^{\rm A} =v\xc^{\rm AE}$, which inputs the instantaneous density into the exact ground-state functional.  Charge begins to transfer but ultimately turns back, even when $v\xc^{\rm A} =v\xc^{\rm AE}$ is used. Panel b) in Fig.~\ref{fig:hubbard} gives an example of this for a very simple system, the asymmetric Hubbard dimer, that nonetheless contains the same physics in this case; the top panels (charge-transfer regime) corresponding to charge-transfer between closed-shell fragments, and the bottom panels (Mott-Hubbard regime) corresponding to charge-transfer between open-shell fragments. . 
As further discussed in the caption of Fig.~\ref{fig:hubbard}, charge-transfer between open-shell fragments of a closed-shell system is instead problematic from the very start, because of static correlation: the KS HOMO and LUMO of the closed-shell system are delocalized over both fragments (like bonding and antibonding orbitals). Semilocal functional approximations yield unphysical fractional charges in the ground-state, while the exact ground-state functional has a step feature that re-aligns the two HOMOs~\cite{M17}. 
We note these concerns do not necessarily apply with such ferocity to other types of charge transfer, e.g. from injection of an electron, or  driven by nuclear motion~\cite{M17}; adiabatic TDDFT with range-separated hybrid functionals tends to be more successful in these cases.

The failure to capture resonant driving in general can be viewed as a consequence of spurious peak-shifting~\cite{RN12}: with an adiabatic approximation, the resonant frequencies of a system that has been driven away from a ground-state spuriously change in time even when that driving field has been turned off. 
Depending on the size of this shift, it can muddy the interpretation of time-resolved spectra in pump-probe experiments, for example, and it has been found to be noticeable in model systems as well as small molecules~\cite{RN12,FLSM15,HTPI14,PI16}. 
Although it has not been explicitly shown, the steps and peaks in the exact xc potential continually bring the otherwise drifting resonant frequencies back to their fixed positions.

\section{Therapies, but not always tried and true}
\label{sec:therapies}
\begin{figure}
\centering
\includegraphics[width=0.8\linewidth,height=0.4\textheight]{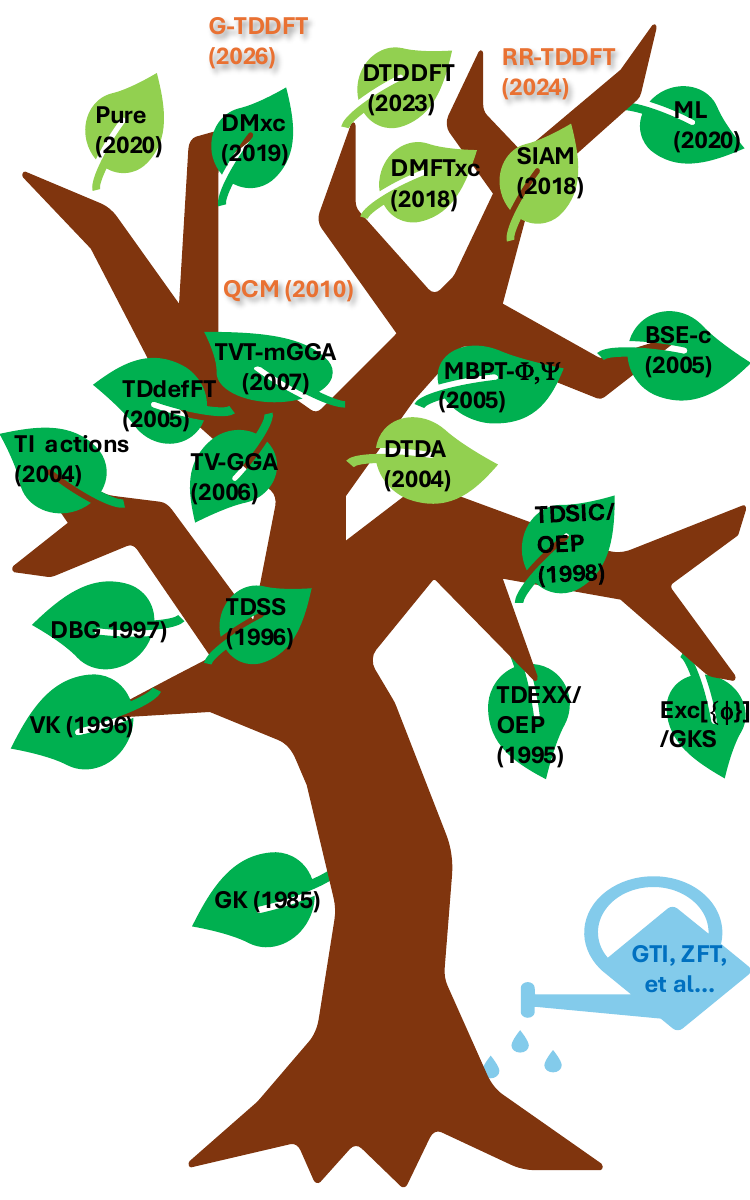}
\caption{Approximations developed for going beyond the adiabatic approximation, rising up from earlier ones to more recent ones as one climbs the tree.  More details are given in Ref.~\cite{LM23}.
The most robust approximations are watered by exact conditions such as GTI and ZFT (see text).
The first, GK~\cite{GK85,GK90} is a frequency-dependent LDA; its violation of GTI and ZFT led to
the idea of local memory following the fluid element in DBG~\cite{DBG97} and a generalization to current-density functionals in TDCDFT~\cite{GD88,VK96} with the VK approximation~\cite{VK96,VUC97,Vignalechap}, with later extensions to a GGA form (TV-GGA)~\cite{TV06} and a meta-GGA (TVT-mGGA)~\cite{TVT07}.
Translationally-invariant (TI) memory actions were developed in Refs.~\cite{KB04,KB06}, and time-dependent deformation functional theory TDdefFT~\cite{TP03,T05,T05b,T07}. On the other side, orbital functionals, with implicit non-local density-dependence, have been explored with TDOEP formalized in Ref.~\cite{UGG95}, including exact exchange (EXX)~\cite{G98,HB08,HB09,HG13}, and self-interaction corrected LDA (SIC)~\cite{URS98,HKK12}. These, and other ground-state energy functionals of orbitals $E\xc[\{\phi\}]$, have been applied also within a generalized KS scheme (GKS)~\cite{SGVML96,GL97,GNGK20}, 
Many-body perturbation theory (MBPT) was mapped to DFT via Sham-Schl\"uter (TDSS)~\cite{L96,TP01}, and through conserving $\Phi$  or $\Psi$-derivable approximations ~\cite{BDLS05}. Relatedly, Bethe-Salpeter-Equation-contracted (BSE-c) xc kernels were found~\cite{RORO02,BSOSR05,STP04,BSSR07,GORT07}; in particular, frequency-dependence arises from contracting over spatial indices~\cite{GORT07}. 
DMxc are density-matrix coupled approaches~\cite{LM19}, and ML represent machine-learning approaches~\cite{SNH20,BCGI22,BI26,WZLZ26} (see text).
Turning to approximations for targetted purposes: DTDA is a dressed Tamm-Dancoff-derived frequency-dependent kernel  capturing double-excitation energies~\cite{MZCB04,CZMB04,MMWA11,C05,M22,HIRC11,CH12,M22} while DTDDFT recovers also their oscillator strengths~\cite{DM23,DM25,DM25Err}. 
The `Pure' functional~\cite{CBR20} contains an implicit frequency-dependence that opens the KS gap in semiconductors and insulators. SIAM derives a non-adiabatic kernel for the Anderson impurity model~\cite{DSH18,DHK19} and DMFTxc for the Hubbard model~\cite{TR17,ATZR20}.
Alternate TD(C)DFT-based approaches float through, capturing non-adiabatic effects in different ways: Quantum continuum mechanics (QCM)~\cite{TGVT09,GTVT10,GJTD12}, RR-TDDFT~\cite{DBM24} (see text) and Geometric-TDDFT~\cite{CDGLLT26,CDGLLT26b} (see text).
}
\label{fig:nonadia}
\end{figure}

Approximations going beyond the adiabatic approximations were proposed and tested soon after the birth of the Runge-Gross theorem, branching out in several directions as illustrated in Figure~\ref{fig:nonadia}. While some have had success for specific types of problems particularly in the linear response regime, a fully non-equilibrium approximation that has essential features of the exact $v\xc$ has so far remained elusive. 
An ideal goal that the universality of the theorems would encourage us to aim for is an all-purpose functional approximation, incorporating known exact conditions associated with spatial and time dependence of the density and dependence on the initial states. At the same time there have been new related approaches describing processes beyond linear response within which the functionals may be more amenable to simple approximations~\cite{CDGLLT26,CDGLLT26b,DBM24}, while at another extreme, machine-learned functionals offer an alternative perspective~\cite{SNH20,BCGI22,BI26}.

Figure~\ref{fig:nonadia} shows the branching of different proposed approximations that go beyond the adiabatic approximation. The caption contains only a short description and some relevant references, and we refer the reader to  a recent review~\cite{LM23}
for a more detailed description of the different approaches that were proposed prior to about 2022. In the following, we go into more detail for the advances made since then. 

Before doing so, we make some general observations. First,  just as in ground-state DFT, exact conditions form a trusty guide for robust approximations. TDDFT inherit some from the ground-state (e.g. self-interaction-free), but there are several known exact conditions without any ground-state counterpart. These have been highlighted in Figure 4 of Ref.~\cite{LM23} and its associated discussion, and include the zero-force theorem (ZFT) and generalized translational invariance (GTI)~\cite{D94,V95,V95b,GDP96},
symmetry and reciprocity relations~\cite{VK98}, causality~\cite{L98,WYB12}, the memory condition intertwining initial-state- and history-dependence~\cite{MBW02,DLM22} (which still plays a role in the structure of the history-dependence even when the initial states are ground states), behavior under uniform coordinate and time scalings~\cite{HPB99}, constant resonant frequencies for a given transition in absence of external time-dependence~\cite{FLSM15,LFM16,HTPI14,RN12c}, and the reduction to the ground-state functionals in the adiabatic limit for systems with a gap~\cite{WYB12}. 
While the severity of the consequences of the violation of these conditions depends on the situation, the particularly egregious ones yield unphysical and unstable predictions, and can be held responsible for failures, e.g. violation of  ZFT leads to self-excitation~\cite{MKLR07,KB08,SLKSU21}, and  spurious peak-shifting leads to failures of resonantly-driven processes~\cite{FLSM15,LFM16,RN12c}.

A second observation is that not all the errors in adiabatic TDDFT predictions are the fault of the adiabatic approximation {\it per se}: sometimes it is that the spatial-dependence on the density is too short-ranged in either or both of the ground-state xc potential or the xc kernel. For examples, for even the lowest excitonic peak in a semiconductor, one needs an xc kernel with $1/q^2$ dependence~\cite{RORO02,BSOSR05,BSU20}, for Rydberg excitations in a molecule  the ground-state xc potential needs a non-local dependence on the ground-state density to yield a $-1/r$ tail~\cite{TAHR99,WMB03}, and long-range charge-transfer excitations require non-local density dependence such as provided by exact-exchange in the ground-state xc functional in both pure- and generalized-KS approaches~\cite{SKB09,K17b,M17}, additionally non-local spatial density-dependence and frequency-dependence in the xc kernel for pure density-functional approaches~\cite{M22,HG12,HG13}.
Looked at from a different angle however, if a functional does have time non-local dependence, errors from inadequate spatial non-locality cannot be completely disentangled from history-dependence~\cite{VK98,Vignalechap}: a time-non-local approximation that is spatially local violates the ZFT, ruling out any local, or even semi-local, density approximation with memory. 

The examples above all fall within linear response, which brings us to our third observation: there is no guarantee that a functional that performs well in the linear response regime performs well for fully non-equilibrium processes simulated by the TDKS equations. For example, while range-separated hybrids can be tuned to accurately reproduce charge-transfer excitation energies~\cite{BLS10,KSRB12, SKB09,KKK13,KB14}, evaluating them on a fully non-equilibrium system introduces severe errors that prevent even a qualitative description of charge-transfer dynamics~\cite{RN11,FERM13,M17,DBM24}.

A fourth observation relates to initial-state dependence. Even when beginning in ground-states which are functionals of the initial density by the Hohenberg-Kohn theorem, initial-state dependence still structures the history-dependence, due to the memory condition~\cite{MBW02}. But the functional-dependence is not on all $3N$ complex degrees of freedom of the initial states, but rather just on the initial interacting one-body reduced density matrix (RDM) and diagonal of the two-body RDM, and the initial one-body RDM of the chosen KS initial state. This can be seen from the exact expression for the time-dependent xc potential~\cite{L99,LM19,LFSEM14}.

\subsection{More Modern Therapies}
\underline{\it{Machine-learned functionals (2020 onwards)}}
Naturally, artificial intelligence explorations have extended to the development of improved memory-dependent functionals. 
Thankfully these efforts have so far remained outside the ``AI empires"~\cite{KarenHao} avoiding the highly problematic operating procedures that exist in these companies at the topmost levels. 
Instead, neural networks have been developed in in-house codes that attempt to learn the xc potential for a given prescribed density evolution. The one-dimensional electron-hydrogen scattering example~\cite{SLWM17,LSWM18} has emerged as a popular and challenging test case, due to the clear evidence of history and initial-state dependence  in yielding accurate dynamics here. 
The problem for $v\xc$ is difficult from the very start when the initial KS state is chosen as a single Slater determinant, since this state has such a different structure than the minimally two configuration interacting state (one electron in a target ground-state and one electron in a moving Gaussian wavepacket).  
The
neural net construction in Ref.\cite{SNH20} learned $v\xc$ from input densities, and 
succeeded in reproducing step and peak features of the correlation potential, showing that the learned functional including memory through a Gaussian-weighted history feature gave better dynamics than a trained functional without memory on related scattering problems not used in the training. 
Instead, by viewing the xc functional as a control that guides TDKS propagation, one can avoid the need for accurate {\it a priori} input data for $v\xc$, an idea that was first demonstrated in Ref.~\cite{BCGI22} and recently extensively studied and expanded  in Ref.~\cite{BI26}, including to beyond one-dimensional models. 
In fact, two pathways are pursued in Ref.~\cite{BI26}:
In one, inversion is carried out first using a 
combination of PDE-solving and adjoint-based optimization to learn the best pointwise $v\xc(\br,t)$ from a given $n(\br,t)$ input, 
without a neural net. The pairs of density-potential functions are then used to a train a neural net in a similar spirit to Ref.~\cite{SNH20}, but the refinement of the $v\xc$ resulting from the first step greatly improves the accuracy of the training data. In a second pathway, the adjoint-based neural-network model combined with the TDKS-constrained optimization is treated to directly learn $v\xc$, avoiding, the inversion step, and it can be used  to learn its density-functional-dependence.
Again their functional generalized well to parameters outside the training set for two-electron scattering problems now in two dimensions. Ref.~\cite{BI26} pointed out the need for high quality numerical propagation methods to generate accurate time-dependent densities as input training data; learning the time-dependent $v\xc$ is very sensitive to this. 
An alternate approach that avoids the TDKS constraint, and in fact avoids time-stepping altogether, but instead encodes a Hamiltonian-consistency is the
``three-head neural architecture''  presented in Ref.~\cite{WZLZ26}. This includes training for symplectic structure together with a memory-dependence as a way to ensure long-time stability, energy conservation, and non-growing errors from discretization. The extent of history-dependence is learned from the data instead of prescribing it. Incorporating Hamilton's equations into the neural network architecture to learn the potential was explored in the classical case in Ref.~\cite{Bhat2019} and had been previously applied to TDKS in Ref.~\cite{YW23} but for predicting adiabatic TDDFT propagation only.
In Ref.~\cite{WZLZ26} the three heads function as a coupled system, with the learning of the symplectic structured dynamics and memory-based potential reinforcing each other. 

\underline{\it{Response-Reformulated TDDFT (2024 onwards)}} Noting that the adiabatic functionals are, by definition, more suited to be evaluated on domains close to a ground-state, and further that
functional developments beyond adiabatic have been far more successful in the linear response regime, Ref.~\cite{DBM24} proposed a ``response-reformulated TDDFT'' (RR-TDDFT) where xc enters only for ground-state, linear and quadratic response quantities, and yet can describe fully non-equilibrium dynamics. The structure of RR-TDDFT is similar to that of time-dependent configuration interaction, in the sense dynamics enters only in propagation of coefficients of fully-interacting wavefunctions, but in RR-TDDFT, these wavefunctions are never found, and the only ingredients in the propagation are the externally applied field and static electronic structure quantities: ground and excited state energies and couplings between the excited states, all of which are formally exactly available from ground-state DFT, linear and quadratic response TDDFT. The key point is that separation of the time-evolution from the electronic structure ingredients means xc functionals are required to be evaluated only near a ground-state, where the adiabatic approximation  performs far better than when these functionals are evaluated on fully non-equilibrium system as they would be in TDKS. Ref.~\cite{DBM24} demonstrated that RR-TDDFT can accurately describe resonantly-driven dynamics, while TDKS completely fails, even when both methods use the same approximate functional. RR-TDDFT can exploit improved functionals developed in the response regime for describing fully non-equilibrium dynamics, as was shown in Ref.~\cite{RBDM26} where dynamics reaching states of double excitation could be reliably predicted, and it can resolve puzzling predictions of the standard TDKS propagation that arise from the incorrect nonlinearity of the xc potential when there is a lack of memory, such as the molecular dipole instability after an ultrafast XUV pulse found in adiabatic TDKS propagation ~\cite{DRM26}.
The scaling depends more heavily on the number of excited states anticipated to be involved in the dynamics, in particular in computing their interstate couplings; whether the method could be applied to extended systems is unclear without further approximations. 

\underline{\it{Geometric TDDFT (2026 onwards)}}
A fundamentally new formulation was recently presented in Ref.~\cite{CDGLLT26,CDGLLT26b}, with new xc concepts and objects, and indications are that their non-adiabatic features are `smaller' than those in the conventional TDDFT. 
The original  formulation of TDDFT could be viewed as resulting from the wavefunction variational principle performed on a constrained manifold of wavefunctions satisfying a target time-dependent density (not to be confused with the variational principle with respect to the density~\cite{L98,V08}), such that the residue of the unconstrained TDSE belongs to the normal space ${\cal N}_{\Psi(t)}$ of possible wave functions that have this target density~\cite{CDGLLT26b}. 
Instead of using the variational principle, geometric TDDFT minimizes the norm of the residue of the unconstrained TDSE for wavefunctions evolving on the manifold of target density, leading to the residue instead belonging to $i{\cal N}_{\Psi(t)}(t)$. 
This results in a Schr\"odinger equation with a {\it complex} but norm-conserving potential, $\hat{W}(t)$. This complex potential can be interpreted as replacing the large peaks and step structures of conventional TDDFT with sources and sinks, locally changing the density to match the target, and a corresponding Geometric Kohn-Sham system can be established. Examples on model systems show that the non-adiabatic components of this complex potential  tend to be much smaller than those of the conventional TDDFT. How to approximate this potential and how it performs on a wide range of non-equilibrium situations remains to be seen but certainly offers a fresh and promising approach.


\section{Outlook}
Progress has been made in understanding where the adiabatic functionals do and do not work, particularly in the linear response regime. Correspondingly, building improved non-adiabatic functionals has so far been simpler in the response regime, as well as improved adiabatic functionals with better spatial-dependence. To illustrate the relevance of such developments, one could consider photochemistry applications, typically modelled by nuclear trajectories or wavepackets moving on coupled electronic potential energy surfaces, with the nuclear rearrangement leading to photo-induced chemical reactions and their control. As highlighted by a recent exciting ``calibration exercise"~\cite{Cyclobutanone}, accurate electronic structure quantities play a key role in the outcome of the excited-state dynamics, more critically so than the method used for the nuclear motion. Given the changing character of electronic states over the landscape of nuclear configurations, improved TDDFT functionals in the linear response are still needed. Further, the RR-TDDFT framework now makes these developments applicable also for simulating fully non-equilibrium processes. 
Fresh approaches such as Geometric-TDDFT and machine-learning discussed here are also promising, and there are certainly other budding ideas being proposed, including using the density's second time-derivative as basic variable~\cite{TU21}, and coupling-constant integral transforms~\cite{G97,LM20b}. Further, the role of the breaking of ${\mathcal C \mathcal T}$-symmetry in approximate non-linear theories brings a new perspective~\cite{FC25}, that may lead to a paradigm shift in the development of new approaches.

Even if a robust approach to xc in all regimes is found, a remaining problem is how to model
functionals for observables not directly related to the density. A particularly relevant example is the current-density: even with the exact xc functional, the KS system in TDDFT only reproduces the longitudinal part of the interacting system's current exactly, and an extra functional would be needed to correct the rotational part of the KS current to that of the true system.  Instead, the KS system of TDCDFT in principle reproduces the rotational component directly~\cite{GD88,V04},  and further allows external magnetic fields. But propagating with the generally spatially-dependent xc vector potential that these calculations require render the computational problem challenging, and to treat non-metallic systems, functionals other than VK and its relatives would be desirable to avoid ~\cite{VK96,TV06,TVT07} unphysical linewidths and decay in systems with a gap~\cite{UB04}.

The task is challenging but also exciting and creative. Given the promise of approximate TDDFT, with its favorable system-size scaling on top of a rigorous foundation,
 it is certainly worth developing further therapies, including improved functionals and alternate related approaches, to build on the results and insights it has already provided.

\section{Funding Information}
This work was partially supported by the U.S. Department of Energy, Office
of Science, Office of Advanced Scientific Computing Research and Office of Basic
Energy Sciences, Scientific Discovery through Advanced Computing (SciDAC) program under Award Number DE-SC0026088.

\section{Acknowledgments}
The author is  grateful to Harish Bhat for very helpful  comments on the manuscript.

\section{Data Availability}
Data sharing is not applicable to this article as no new data were created or analyzed in this study.

\bibliography{ref.bib}
\end{document}